# "Ideal-Chain Collapse" in Biopolymers

Richard M. Neumann

The Energy Institute
The Pennsylvania State University
University Park, PA 16802

(12/19/00)

## Abstract

A conceptual difficulty in the Hooke's-law description of ideal Gaussian polymer-chain elasticity is sometimes apparent in analyses of experimental data or in physical models designed to simulate the behavior of biopolymers. The problem, the tendency of a chain to collapse in the absence of external forces, is examined in the following examples: DNA-stretching experiments, gel electrophoresis, and protein folding. We demonstrate that the application of a statistical-mechanically derived repulsive force, acting between the chain ends, whose magnitude is proportional to the absolute temperature and inversely proportional to the scalar end separation removes this difficulty.

## Introduction

For nearly 60 years, Hooke's law has provided a convenient means of describing the entropic elasticity of an *ideal* Gaussian polymer chain. [1] Frequently, the force law for a single chain is given by

$$f_h = -(3k_B T/nl^2)r, \qquad (1)$$

where $f_h$ is the average attractive force acting between the chain ends separated by a scalar distance $r$. $k_B$ is the Boltzmann constant, $T$ the absolute temperature, $n$ the number of chain links, and $l$ the link length. It is apparent that in the absence of an external stretching force, such an equation, if interpreted as a macroscopic equation of state, predicts the collapse of the chain because a retractive force is present for $r > 0$. In fact, this interpretation has prompted one popular reference [2] to state "Therefore, we cannot bring in anything like the relative deformation $\Delta r/r$ which appears in the usual form of Hooke's law". This statement appears in the context of defining an elastic modulus to be the ratio of the stress to strain, where the strain, by definition, would require division by zero; zero being the magnitude of $r$ associated with the absence of an external stretching force. Although this paradox is frequently ignored, some authors have attempted to deal with it by focussing solely on real chains and relying on excluded volume to prevent chain collapse. [3] Others have invoked a pair of "Maxwell demons", one at each end of the chain, pulling on it with a force equal to the retractive force. [4] Still others have suggested that the origin of the problem lies in the microscopic nature of the system, resulting in a nonequivalence among ensembles derived by holding either the length, force, or displacement constant. [5,6]

Here, in the spirit of the latter view, we show that the following thermodynamic relationship, which expresses the average repulsive force acting between any pair of particles as a function of their separation-distance $r$,

$$f_r = 3k_B T/r, \qquad (2)$$

provides a means of avoiding the paradoxical behavior noted. [7] In analyzing DNA-stretching data, Eq. 2 used in conjunction with Eq. 1 predicts a chain end separation, at zero stretching force, equal to the root-mean-square value, $r_o$ ($r_o = n^{1/2}l$), in agreement with experiment [8], rather than the zero result predicted by Eq. 1 alone, used by the authors [8] to interpret their results. In the lakes-straits model for polyelectrolyte migration in gel electrophoresis, the chain-collapse problem is circumvented by the authors through the derivation of a repulsive force proportional to $T/n$. [9] We demonstrate that in the absence of an external field, such a force will move chain segments from a pore having few segments to an adjoining pore having many, in apparent violation of the Second law. We suggest that Eq. 2 provides a statistical-thermodynamic remedy for this problem. In a recent model of protein folding based on springs and beads [10], the excluded volume of the beads is used by the authors to prevent the collapse of the protein into a point singularity. Their approach leads to an equilibrium bead separation that is not dependent on the temperature. We demonstrate, using Eq. 2, that the classical dependence of the end separation of a spring on the temperature [11] is recovered and that excluded volume is not required to prevent the collapse of a spring.

**Theory**

Whereas there are a number of ways to derive Eq. 2 [7], a particularly simple and instructive approach is to examine a Brownian particle located inside a spherical volume of radius $r$. The particle occupies a volume element equal to $4\pi r^3/3$, and the particle concentration, $c$, is proportional to $r^{-3}$. A chemical potential may be defined by $\mu = k_B T \ln(c)$, permitting the derivation of Eq. 2 using $f_r = -(\partial \mu / \partial r)_T$. $f_r$ may be viewed as the *average* repulsive force acting between two particles separated by a distance $r$, one particle being located at the center of a spherical-coordinate system, $r = 0$. Equations 1 and 2 may be combined to yield a force equation for an ideal Gaussian chain for which the force vanishes when $r = r_o$, rather than when $r = 0$:

$$f = 3k_B T(1/r - r/r_o^2) = m(1/h - h), \tag{3}$$

where $h = r/r_o$, and $m = 3k_B T/r_o$. Unlike Eq. 1, which does not permit the definition of an elastic modulus in the usual sense, Eq. 3 yields an elastic modulus equal to $2m$.

**Results and Discussion**

*DNA-Stretching Experiment*

Figure 1 shows the extension of a DNA molecule, visualized with fluorescence microscopy, as a function of force derived from the flowing solvent surrounding it. [8] The figure depicts only the qualitative features of the experiment in the weak-stretching region. The molecule was held stationary against the flow by means of a microsphere attached to one end; the microsphere, in turn, was secured by means of optical trapping. The measure of extension used by the authors is the average maximum visual elongation in the direction of flow, which for our purposes may be approximated by $r$. The straight line depicts an analysis based on Eq. 1 used by the authors. The discrepancy between the data and theory is obvious, most notably the fact that the extension measured at zero flow rate is not zero. Equation 3 is consistent with the experimental data in that at zero external force, $r$ approximates the random-coil value, $h = 1$; with increasing force, Hooke's law (Eq. 1) is recovered because $f_r$ becomes insignificant for large values of $r$.

*Lakes-Straits Model in Gel Electrophoresis*

Figure 2 shows a model for a gel where the pores are represented by "lakes", and the straits are the narrow regions connecting the lakes, where there is just sufficient room through which an idealized DNA molecule can migrate. The chain is sufficiently long so that its various sections reside in a series of lakes. In describing the elastic behavior of the migrating chain, the authors [9] derive a force equation for an arbitrary section of chain using the distribution function for an ideal Gaussian chain in a given lake of size $r$ (the strait-separation $r$ coincides with the end-to-end separation of the chain section), $P(r,n) \approx n^{-3/2}\exp(-3r^2/2nl^2)$. The entropy is calculated in the usual manner via the Boltzmann expression, $S = k \ln(P)$. Here, the lake size is regarded as fixed, the number of links, $n$, present in the pore fluctuates, and $S$ is differentiated with respect to $n$ (rather than $r$) to obtain the force equation,

$$f_n = (3k_BT/2nl)(1 - r^2/nl^2). \qquad (4)$$

As in Eq. 3, a positive value for $f_n$ indicates a repulsive force; a negative value indicates an attractive force. It is readily apparent that the force vanishes for $nl^2 = r^2$; i.e., when the number of chain links is equal to the value for a random coil having an end separation equal to $r$. Thus, it would appear that the paradox of chain collapse has been avoided. It is instructive, nevertheless, to consider the following example where two adjacent lakes ($i$ and $j$), connected by a strait, are prepared so that $n_j > n_i > (r/l)^2$. Equation 4 predicts that, in the absence of an applied field, segments will migrate from lake $i$ to lake $j$, in apparent violation of the Second law. The reason for this behavior is subtle and based on the implicit use of a microcanonical-ensemble in deriving Eq. 4; the reader should consult the references for additional background. [9] We simply suggest that for calculations or simulations of electrophoretic behavior based on the lakes-straits model, Eq. 3 rather than Eq. 4 should be used to calculate the entropic-elasticity contribution to the equation of motion.

*Gaussian Model of Protein Folding*

Here a protein is modeled by beads and springs wherein all the interactions between any given pair of beads are governed by a single quadratic potential. [10] The beads represent monomers that may be polar or hydrophobic, and the springs represent the covalent and noncovalent attractive forces acting between the beads. The model is Gaussian because of the use of a quadratic potential that mimics the Hookean behavior of an ideal Gaussian polymer-chain network. The authors note that, as in the classical theory of rubber elasticity, the springs in their protein model are prone to collapse [3,10], regardless of the temperature. Thus, they include a repulsive potential whose magnitude is adjusted to provide the various stable conformations of the protein being modeled. Because the calculations for an entire protein require the use of a matrix method to describe the folding (relaxation) process and the stable states and because we are concerned solely with the behavior of a given pair of beads and their connecting spring (where the spring energy is of order $k_BT$) *in a heat bath*, we shall consider the simplest system described by the model: two beads attached to each other by a spring – a Hookean dumbbell.

Following the authors' approach for the multi-bead system, we shall solve the Langevin equation for the dumbbell,

$$-\zeta d\mathbf{r}/dt + \mathbf{f}(t) - \partial U/\partial \mathbf{r} = 0, \qquad (5)$$

where $\mathbf{r}$ is the vector separation between the first bead and the origin of a coordinate system located on the second bead, $\zeta$ is the friction factor for a bead, $\mathbf{f}(t)$ is the random-force vector, and $U$ is a potential, here equal to $ar^2/2$. With the use of a coarse-grained time scale to permit $\mathbf{f}(t)$ to average to zero, the relaxation expression resulting from the solution of Eq. 5 is $\mathbf{r}(t) = \mathbf{r}_o\exp(-t/\tau)$. Note that the relaxation time $\tau$ is independent of temperature; $\tau = \zeta/a$. Thus, at long times (equilibrium) and in the absence of excluded volume, $\mathbf{r} = r = 0$. However, in the classical treatment of Brownian motion by Chandrasekhar [11], the equilibrium separation is given by,

$$<r^2> = 3k_BT/a. \qquad (6)$$

This result indicates that it is inappropriate to average out $\mathbf{f}(t)$, which is the only term in Eq. 5 that reflects the thermal energy in the spring-bead system through the relationship $<\mathbf{f}(t_1)\mathbf{f}(t_2)> = (2\zeta k_BT)\delta(t_1 - t_2)$ [11].

Equation 2 provides an alternate, and very much simpler derivation of Eq. 6; the repulsive force $3k_BT/r$ is simply equated to the spring-force $ar$ to yield $r^2 = 3k_BT/a$. Thus, in the Gaussian model of protein folding, the expressions for the equilibrium average separation between pairs of beads must include terms proportional to $k_B T/a$ *in addition* to the contributions from whatever excluded volume is present. The authors claim to present a "thermodynamic" (and presumably statistical mechanical) approach to protein folding but, in fact, have obtained a mechanical, nonthermal result because of their neglect of $f(t)$ in solving the Langevin equation. In other words, the temperature does not appear in their results.

## Conclusion

Equation 2 is essentially a manifestation of the ideal gas law for a microscopic system consisting of one particle. Equation 3 was anticipated by Flory [12] in his derivation of the molecular expansion factor, $\alpha$; however, most of his work on rubber-like elasticity is based on Hooke's law (Eq. 1). Using Eq. 2, we have presented realistic solutions to three different problems, all associated with a polymer chain (or spring) in a heat bath.

It is obvious that the conventional Hooke's-law approach cannot provide an adequate description for polymer-stretching experiments in the weak-force regime where the chain's elongation is measured as the *scalar* end-to-end separation. Equation 3 provides a description in qualitative agreement with experiment in that it yields the random-coil end separation at zero applied force; in the intermediate-force regime, it reduces to the Hookean force law.

Whereas the lakes-straits model force equation (Eq. 4) describes an ideal chain that does not collapse in the absence of an external field, it does predict spontaneous segmental flow from a region of low density to one of high density. The present approach using Eq. 3 avoids both the reverse-flow and the chain-collapse problems.

Finally, the Gaussian model of protein folding, despite the claim of its authors that it is a thermodynamic model, does not incorporate the effect of thermal motion. This occurs because the random-force vector in the Langevin equation is averaged out, resulting in dynamical solutions and equilibrium-state configurations characteristic of a mechanical, nonthermal system. Applying Eq. 2 to a simple bead-spring dumbbell in a heat bath, we obtain the classical result for the equilibrium size of a harmonic oscillator and suggest that the protein-folding model requires terms proportional to $k_B T/a$ in its expressions for the equilibrium separations of its beads.

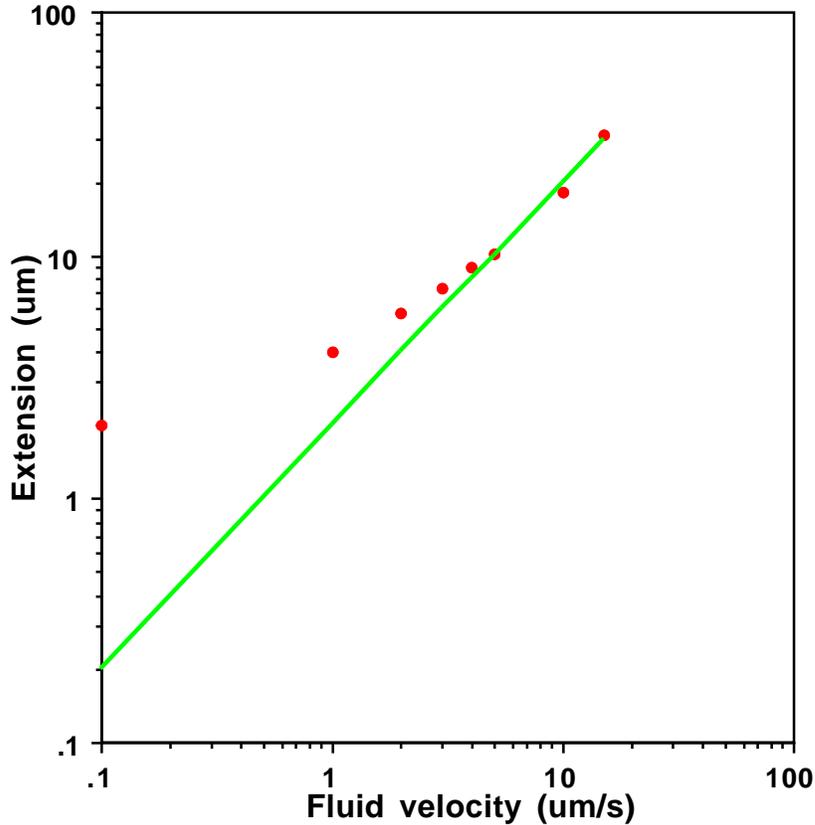

**Figure 1** (•) depict extension measurements as a function of the fluid velocity for an individual, fluorescently labeled DNA molecule using data in the weak/moderate-stretching regime adapted from Fig. 1 of Ref. 8. The initial extension here is based on an estimate for $r_o$, as the authors provided no value for this quantity. The solid line (—), also borrowed from Ref. 8, illustrates Hookean behavior "predicted" from a dumbbell model based on Eq. 1. In the experiment, a series of "maximum visual extensions in the direction of flow" for a single fluctuating molecule (at a fixed solvent flow rate in the $x$ direction) was measured. The individual measurements, taken at approximately 1-second intervals, were averaged to obtain the reported extension, which corresponds to $<|x|>$ rather than $<r>$. Because $<|x|> \rightarrow <r>$ with increasing force, a distinction between these two measures of extension is not necessary for the present purpose.

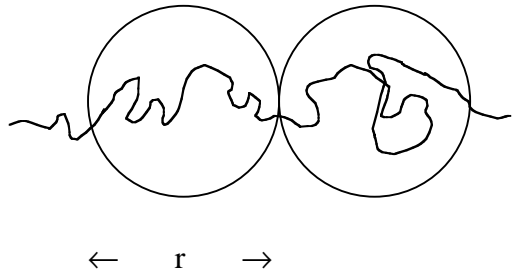

← r →

Figure 2. Two adjoining lakes are shown, each of size *r*. The polymer chain is depicted as the "scribble" line passing through both; the strait is the point of contact between the two circles.